# Structure, stability and elasticity of DNA nanotube

Himanshu Joshi[1], Anjan Dwaraknath[1,2] and Prabal K Maiti[1*]

[1]Centre for Condensed Matter Theory, Department of Physics, Indian Institute of Science, Bangalore India 560 012

[2] Indian Institute of Technology Madras, India 600 036

## Abstract

DNA nanotubes are tubular structures composed of DNA crossover molecules. We present a bottom up approach for construction and characterization of these structures. Various possible topologies of nanotubes are constructed such as 6-helix, 8-helix and tri-tubes with different sequences and lengths. We have used fully atomistic molecular dynamics simulations to study the structure, stability and elasticity of these structures. Several nanosecond long MD simulations give the microscopic details about DNA nanotubes. Based on the structural analysis of simulation data, we show that 6-helix nanotubes are stable and maintain their tubular structure; while 8-helix nanotubes are flattened to stabilize themselves. We also comment on the sequence dependence and effect of overhangs. These structures are approximately four times more rigid having stretch modulus of ~4000 pN compared to the stretch modulus of 1000 pN of DNA double helix molecule of same length and sequence. The stretch moduli of these nanotubes are also three times larger than those of PX/JX crossover DNA molecules which have stretch modulus in the range of 1500-2000 pN. The calculated persistence length is in the range of few microns which is close to the reported experimental results on certain class of DNA nanotubes.

[*] Email address: maiti@physics.iisc.ernet.in

**Introduction:**

DNA's structural and complimentary base pairing properties[1] makes it an attractive molecule for nanotechnology. DNA as a nanostructure building block was first identified by Nadrian Seeman in the 1980s. Since then many different DNA motifs have been designed. Many of these motifs were made of simple elements, such as sticky ends and DNA crossover molecules[2]. Sticky ends are single stranded DNA tails on bigger molecules which can be hybridized with another such complementary tail on another molecule, thus joining the two molecules of DNA. Larger structures can thus be created with controlled assembly. This led to a variety of motifs which were topologically similar to cubes[3], truncated octahedrons[4] and other polyhedral[5-7]. But their actual microscopic structures are not known accurately as the double helix DNA molecule does not provide enough rigidity to make 3-dimensional rigid nanostructures. In order to make rigid structures out of DNA, the crossover DNA concept was introduced[2,8-10]. In the crossover DNA structure one DNA strand exchanges itself with another strand, from one domain to a parallel domain, thus linking the two domains together to form a Holliday like junction. The DX molecule is made by having two crossovers between two domains and thus resulting in a rigid structure with sticky ends at both ends of each domain. This can be used as a versatile building block. Many structures were made using the DX molecules, like self-assembled sheets such as DX arrays[11]. The DX molecules can also be used to make patterns on these sheets by algorithmic self-assembly. This involves a careful selection of sticky ends which gives precise control over the assembly, thus making striking patterns such as the Sierpinski triangle at a nano scale[12]. The DX molecules were used to make nanotubes as well[13]. The DX molecules were further modified to form other motifs such as the PX and JX structures which were used to make a robust nano machine.[14,15] This opened up new area, with many possibilities by using the simple PX-JX2 machine as a building block for more complicated nano machines. The thermodynamic stability of these structures was also studied by fully atomistic MD simulations[16,17]. Since the actual structure cannot be determined by experiments, computer simulation is used to model these structures to get a comprehensive relationship between the structure, stability and their elasticity.

Experimentally DNA nanotubes have been constructed by various ways. They can be constructed by using the DX molecules to form arrays which wrap on to themselves forming a tubular structure.[13,18] Here the DX molecules function as tiles with specific sticky ends so that there is twist between successive molecules, thus curving the sheet to form a tube. Such nanotubes have variable diameters as the number of DX molecules needed to close the structure can be varied to produce tubes of different diameters. The cross section of such nanotubes maybe skewed.

Another way in which DNA nanotubes are constructed is by using a fixed number of DNA double helical domains to form a closed polygonal cross section. Each domain is attached to its neighbours by two crossover points. By controlling the position of the crossover points one can set the angle formed between these domains. These angles can be set to produce DNA nanotubes of various cross section. In this fashion both 6-helix tube and 8-helix tube have been produced.[19,20] These tubes are first constructed as half tubes and finally two halves are combined to complete the tube.

DNA nanotubes can also be constructed in another way. Square or triangular, horizontal planar templates are first made from DNA double helices along with an appropriate vertex molecule. These are designed to have sticky ends on the vertical axis at each vertex, which enable it to be stacked one on top of each other like the rungs of a ladder. Thus these templates along with other vertical strands can be made into a nanotube. The space in between two such planar templates can be used to store some sort of cargo. The nanotube can be made such that the cargo can be released when a particular strand is introduced. The introduced strand hybridizes with a particular strand on the vertical domains of the nanotube making it single stranded in those regions. This makes those regions more flexible thus allowing its contents to be released.

Although several groups have synthesized DNA nanotubes[13,18-23] the microscopic picture of these systems is not yet clearly known. The experimental fabrication of these structures is intricate and time taking. Here we present a computational algorithm to build these structures and then study the structure and stability of these systems through fully atomistic molecular dynamics (MD) simulations. Non-Denaturing gel electrophoresis and other experimental methods have shown the stability of these structures, but to make devices for nano-engineering, structural, mechanical and electronic properties play crucial role. We have attempted to explore the structural and mechanical properties of these nucleic acid bundles for the first time using fully atomistic MD simulations. We hope that the present work will help in better understanding the microscopic structure of DNA nanotubes which will help for their better and proper usage in nano-engineering. In constant velocity ensemble steered molecular dynamics (SMD) simulation, we also pull these structures and calculate the force extension behavior of these DNA nanotubes. From the elastic regime of the force-extension curve using Hooke's law, we calculate the stretch modulus. Previously, we have used similar technique to study the stretch modulus of PX and JX crossover structures. We give a quantitative measurement of stretch modulus of 6-helix and 8-helix structures through atomistic MD. Using the calculated stretch modulus, we also report the persistence length of these tubes and found them to be in close agreement with the available experimental results.

**Construction and Simulation methodology:**

**Building Protocol:**

DNA nanotubes can be constructed by arrangement of individual DNA double helix molecules in a hexagonal pattern as shown in figure S1 in the supplementary information. Each DNA double helix is connected to the adjacent one by crossovers between adjacent strands. The position of the crossovers in the nucleotide sequence is chosen carefully so that crossovers to different adjacent DNA helices form an appropriate angle between them to close the tube. For example, to get an angle of about 120 degrees corresponding to hexagonal arrangement, the crossovers must be separated by 7 base pairs (figure S1 in the supplementary information.)

We have built the following topologies of DNA nanotubes as shown in figure 1:
1. 6-helix Bundle
2. 8-helix Bundle
3. Triangular nanotubes (Tri-tube)

6-helix and 8-helix nanotubes are composed of half tubes so that they can encapsulate the cargo inside them. A 6-helix nanotube would involve a half tube comprising of 3 double helix molecules attached by crossovers. This molecule resembles triple crossover[24](TX) molecule except that it is bent with an angle of 120 degrees between its outer helices. The TX molecule is effectively the strand-switching among three double helical domains. Two such molecules can come face to face if required enclosing some other long molecule forming a nanotube.

The nucleotide sequences in each double helix forming the nanotube and the positions of the crossovers are taken from the experimental design. We have developed a code that generates DNA nanotube structures of various topologies. The program is written in NAB[25], a programming language which is available in the AMBER software package[26-28]. The details of the construction algorithm are given in supplementary information. The 3-D coordinates from our code is output in PDB format, which can be opened by xLEaP module of the AMBER software package, to generate other files required for simulation. The 6-helix bundles generated by the code form a nice tubular structure as its theoretical angles were close enough to form a hexagonal cross section. The 8-helix bundles on the other hand have very open structure and did not close if arranged according to the above protocol. We need to generate some larger bond to close the 8-Helix bundles. (Figure S5 of supplementary information)

Triangular tubes are constructed using triangular DNA templates which are assembled one on top of each other to form a ladder-like nanotube. This template requires a special corner molecule, TBZ, to connect each DNA side to other in order to form a triangular tube. The structure of the corner molecule is shown in **Fig. S2** in the supplementary information. The construction protocol for this type of nanotube begins with the three scaffold helices: these are long DNA duplex molecules which have open bonds for the triangle rungs to be bonded to. The three scaffolds DNA duplex are positioned first to form a triangle. The DNA rungs are then oriented one by one into place. Finally, the corner molecule is added to connect the DNA sides to form a closed triangle. This involves forming a cyclic strand, which is unsupported by the PDB format that is output from the NAB program. Therefore a script is also generated which when run in xLEap fixes the bonds to form the cyclic strand.

**Simulation details:**
We have used AMBER MD suite of program[26-28] with parmbsc0[29] refined amber force field[30] for DNA and the TIP3P[31] model for water. These force-fields for B-DNA has been validated by previous MD simulations[32]. The structures built from NAB were solvated with water box of using the xLEaP module of AMBER. Some water molecules were replaced by $Na^+$ counterions to neutralize the negative charge of sugar phosphate backbone of DNA double helices. We have used the recent ion parameter from Joung and Cheatham.[33] The LEaP module works by constructing columbic potential on a grid of 1 Å resolution and then placing ions one at a time at the highest electrostatic potential. Once the placement of all the ions is done using the previous method, long MD simulation ensures that they sample all the available space around DNA. We have used periodic boundary in all three directions during simulation. Comparative studies of 3 types of topologies viz., 6- helix, 8- helix and tri-tubes in various combinations of sequences like AT rich, GC rich and original sequence used in experiments have been done. The details of the simulated system are shown in Table 1. After building the structure as discussed above, we simulated the system with standard minimization protocol which was suitable for large DNA nanostructures. Minimization is performed so that system eliminates bad contacts with solvents and ions. During this minimization the DNA nanostructures were kept fixed in their starting conformations using harmonic constraints with a force constant of 500 kcal/mol/Å$^2$. This was followed by series of conjugate gradient minimization while decreasing the force constant of the harmonic restraints from 500 kcal/mol/ Å$^2$ to zero in steps of 5 kcal/mol-Å$^2$. The minimized structures were then subjected to 40 ps NPT (P=1 atm, T=300K) MD using 1fs time step for integration. During this period systems were heated gradually from 0 to 300 K using a 20 kcal/mol-Å$^2$ harmonic restraint on solute to its initial structure. During dynamics, all covalent bonds involving hydrogen atoms were constrained using SHAKE algorithm[34]. After this equilibration, the system undergoes 100 ps NPT dynamics with 2 fs time step to achieve correct solvent density. Particle Mesh Ewald (PME) method was used to compute the non-bonded electrostatic interaction. Finally we have carried out 50 ns long NVT MD in explicit water with 2 fs time step at 300 K using the Berendsen weak coupling method. We save the trajectories for analysis after every 1 ps. Similar simulation protocol was found to produce stable MD trajectory for various DNA nanostructures.[16,17,35]. To study the mechanical behavior under external force, we apply stretching force on both ends of the DNA nanotubes in constant-velocity ensemble. By doing so we have tried to mimic the nano-manipulation techniques such as AFM, magnetic tweezers or optical tweezers using our atomistic MD simulations. Before pulling we choose wider water box to ensure at least 10 Å solvation shell around nanotubes in the fully stretched form. This makes these simulations reliable but the computational cost increases enormously. After performing 1 ns NPT dynamics we pull them with constant velocity 1 Å/ns (or 0.1m/s). First we identify the last residue of each strands, and then we pull O3' atoms at both ends in the outward direction along the tube length. During pulling, we kept track of the extension of the nanotube end-to-end as a function of the applied force. The stretch modulus is calculated from the linear region of stress vs. strain plot. We have followed the same pulling protocol to calculate the stretch modulus for 38 base-pair B-DNA which comes out to be 967 pN (Supplementary info. **Fig. S7**) and are in good agreement with the experimental and previous simulation results[36,37]. Based on this verification, we expect that this protocol will be valid for DNA nanotubes studied in this paper.

**Results and Discussion:**

**RMSD Analysis**:
To understand the stability of various DNA nanotube structures we have calculated root mean square deviation (RMSD) with respect to initially minimized structures. **Fig. 2** gives the RMSD for 6-helix and 8-helix structures. The RMSD is around 20 Å for 6-helix structures or 3-4 Å per helical domain. Apart from some initial fluctuation, the RMSD of 6-helix AT rich structures show better stability compared to original 6-helix sequence used in experiment. To check the effect of ionic FF on the stability of the nanotube structures we have also done simulations using Aqvist ionic FF[38]. Using Aqvist set of ion parameters, we get a different stability pattern as shown in figure S4 in the supplementary materials. There are reports that JC ion parameter gives rise higher stiffness of the DNA as compared to the Aqvist ion parameters[39] due to higher binding with phosphates. This may give rise to different stability pattern as seen in our simulation..

Notwithstanding the different stability pattern depending on the ionic FF, our simulation results demonstrate that the 6-helix DNA nanotubes are stable tubular structure. The VMD[40] visualization of trajectories (Supplementary information video **SV1**) and instantaneous snapshots of various 6-helix structures at different time interval, shown in **Fig. 3,** also show that base stacking and hydrogen bonding are well maintained and the tubular structure is preserved over several ns long dynamics. To have a quantitative estimate of the hydrogen bonds, in **Fig. 4 (a),** we show the time evolution of the percentage of broken hydrogen bonds for various 6-helix tubes. We find that most of the hydrogen bonds are maintained again demonstrating the stability of these structures. In contrast, the 8-helix structures are quit open. Because of its geometry, it is difficult to make a close packed structure with B-DNA and when forced to produce a closed tubular structure, it results in lot of dihedral strain in the tube structure. It is also reflected in snapshots shown in **Fig. 3.** 6-helix structures maintain the tubular structure nicely but 8-helix structures are highly distorted and more than 25-30% hydrogen bonds are broken as shown in **Fig 4(b).** The snapshots of 8-helix structures show large distortion from the regular tubular structure due to high strain in the helices of these structures. The RMSD analysis for 8-helix structures is nicely flattened. The simulation results confirmed the proposition that the 8-helix nanotube was too strained to form a regular tubular structure. It instead stabilized (as evidenced by the flattening of the RMSD graph) into distorted tubular structure which minimized the excessive strain of its initial configuration. This is supported also by the erratic nature of the radius profile of the 8-helix tubes (**Fig.S6** in the supplementary info). In **Fig 5 (a)**, we have plotted the RMSD of various triangular nanotubes. Tri-tube made of GC sequence shows most fluctuation, but triangular topology remains stable throughout several nanosecond long dynamics. AT rich structure shows better stability compared to the tri-tube structure made of only AT sequence. In the original experiment tri-tube was synthesized with several units stacked on top of each other. To test, if multiple units can better stabilize the tri-tube structure, we have also simulated 2 unit of GC rich tri-tube. **Fig 5 (b)** shows the RMSD of 2 unit GC triangular DNA nanotube. Thus we see that different sequences have varying stability for various topologies. The AT rich structure is better stabilized in 6-helix topology but in 8-helix, it's the original structure which is having very less RMSD fluctuation. Using single stranded overhangs, we can join the nanotubes and get the experimentally realized longer nanotube structure. However, at the moment it will be a computational challenge to simulate such longer nanotube fragment. Recently developed meso DNA model[41-43] may allow us to study such longer DNA nanotube structures in near future. We expect that there will be lesser fluctuation (in terms of RMSD and H-bond analysis) in the average properties of these nanotubes as we increase the length of the nanotubes. The nanotubes will have more regular and flatter radius profile as the end effects decrease as we increase the length of the nanotube.

**Radius Analysis:**

One of the principal applications envisaged for these nanotubes is that they can carry cargo in their cavities for various bio-medical applications. Such application requires a detailed knowledge about the radius of nanotube pore and precise control over that. However, this microscopic structural detail is not available from the available experimental results. So it is very important to have a quantitative estimate of their radius. The radius profile is computed by sampling the structure at 0.5 nm intervals along the nanotube long axis. Each section of 0.5 nm represents a ring whose radius we wish to compute. The center of the ring is computed by taking the average location of each atom in the section. The radius of the section is computed by calculating the root mean squared distance from that center. **Fig. 6** shows the radius for 6-helix DNA nanotubes. On horizontal axis we have plotted the length from the center of the tube while on vertical axis the average radius from the last 10 ns long simulation trajectories. All 6-helix structures show almost same radius profile and have radius close to 2.5 nm at the center and on both sides of the 6-helix bundle, we see larger radius due to finite size of the helix. These structures maintain the tubular form throughout the simulation. In contrast, 8-helix structures are very open structure and it is difficult to define radius profile for them. So the radius profile of 8-helix structures (**Fig. S6** in the supplementary information) are quite zigzag and have radius around 3.7 nm at center. It fluctuates between 3.7 to 4.6 nm away from center, on the both sides. In **Fig.7**, we show the radius profile for triangular tubes. Triangular tubes also show the variable diameter along the length as seen in experiment[21]. The radius varies between 1.5 to 3.5 nm in going from the narrow to wider region. The radius profile for this single unit is very noisy implying less stability due to fewer rungs in the tri-tube structures. To check the stability as a function of the number of rungs in the tri-tube structure we have also simulated tri-tube structure having three rungs. The instantaneous snapshots of the GC tri-tube with two and three rungs have been shown in figure 3 (c) and 3(d) respectively. The tri-tube structure with 3 rungs is better stabilized as is evident both from the RMSD analysis as well as radius profile. Quantitative estimate of the radius profile is one of the important outcomes from our all atom simulation.

**Force Extension Behavior:**

Single molecule experiments have been used to study the force-extension behavior of dsDNA [37,44-46]. In the low force regime the elasticity is dominated by entropy and described well by standard worm-like chain (WLC) model or its other variants[44,45,47]. When the dsDNA is pulled beyond the elastic region, the structure elongates 1.7 times its initial contour length which gives rise to the plateau regime[37] in the force-extension curve. This large elongation with a small change in force can be viewed as either force induced DNA melting[36,48-52] or B-S DNA transition [37,53,54]. WLC model and other available model can't account for the entropic elasticity and explain the experimental observation of the plateau region.[37,53,55] Recently we have shown that similar plateau can be obtained when PX/JX DNA molecules are pulled in all atom MD simulation[35]. We have carried out pulling in steered MD simulation for DNA nanotubes in constant velocity ensemble. (In **Fig. S8 of** the supplementary information we provide the instantaneous snapshots of the DNA nanotube at various extensions. Supplementary Video V3 illustrates the trajectory of the 6-helix AT rich structure in our SMD simulation.) **Fig.8** shows the force-extension behavior for various DNA nanotubes. The force-extension curve consists of entropic region where the extension is about 10% of original length of nanotubes. It corresponds to a force of 400pN, where all hydrogen bonds are intact. Beyond this force, we find the overstretching plateau region. Force-extension curves for 6-helix nanotubes made of all AT and all GC base sequences, show almost similar behavior (**Fig. S10 (c)** in the supplementary information). We also calculate the stretch modulus from the linear region of the force-extension plot using Hooke's law, where strain = $\Delta l/l$ and $l$ = initial equilibrated length. The calculated stretch moduli are 4468 (±270), pN  4270 (± 249) pN, 4507 (± 213) pN and 4397 (± 216) pN for GC rich, AT rich, all GC and all AT 6-helix nanotubes respectively. Because of 3 hydrogen bonds, GC rich structure shows slightly higher stretch modulus compared to AT rich structure. The stretch modulus for all AT and all GC 8-helix structures are 3825 (± 224) pN and 3898 (± 191) pN respectively. Higher strains in the 8-helix structures make them less stable and are reflected in the lower value of stretch modulus of these structures. While applying same methodology for 38-mer dsDNA, we found its stretch modulus to be 967 pN. Earlier we have shown that the PX and JX crossover molecule has stretch modulus of order of 1500 pN . For the 6-helix and 8-helix nanotubes simulated in this work, there are two crossovers per strands in these structures. So the 6-helix and 8-helix nanotube structures have stretch moduli which are 2-3 times higher than those of PX and JX crossover structure. The pulling rate used in our SMD simulation is 1 Å/ns which is very high compared to experimental pulling rates (~ μm/s). In order to see the pulling rate dependency on the elastic response on these tubes,  6-helix AT rich structure has been pulled with 3 different velocities respectively 1 Å/ns, 0.5 Å/ns and 2 Å/ns (Supplementary info S11). The plateau region i.e. helix to ladder transition, starts at lower force regime as we go to slower pulling rate. This is expected to go to ~ 100 pN at experimental pulling rates. We find that the slope of linear region is very similar for all the three pulling velocities and so stretch modulus does not depend on the pulling velocity very much. It also ensures that the hydrostatic resistance is not significant. We also perform WHAM[56,57] analysis to calculate the free energy of these nanotubes from the pulling simulations (**Fig. S9** in the supplementary information). From the minima of the free energy as a function of the DNA nanotube length we compute the effective equilibrium length of these structures. 6-helix AT rich and GC rich and 8-helix AT and GC tubes have effective equilibrium length in the range of 18.4nm, 17.6nm, 17.7 nm and 16.8 nm respectively. The equilibrium length of 6-Helix structure made of all AT and all GC nanostructures are also of similar magnitude.

The persistence length of these nanostructures has also been calculated using the force extension data. The persistence length $L_p$, has been calculated using the formula $L_p = \dfrac{EI}{k_B T}$, where $E$ is the Young modulus of the nanotube, $I$ is the area moment of inertia of the nanotube, $k_B$ is the Boltzmann constant and $T$ is the temperature. We estimate the Young modulus from the stretch modulus (S) assuming the nanotube as a cylindrical object of radius r, giving $E = \dfrac{S}{\pi r^2}$. Using the stretch modulus calculated from the force-extension and radius from the radius profile we have calculated the persistence length for various 6-helix and 8-helix nanotubes. **Table 2** gives the values for the persistence length of various nanostructures. The persistence lengths of the 6-helix and 8-helix bundles are of the order of 6 μm and 7 μm respectively. The calculated persistence length is in quantitative agreement with the available experimental results for 6-helix bundle[58]. So our simulations have predicted quite accurate estimate of the stretch modulus as well as the radius of these DNA nanotube structures.

**Effects of Sequence and Overhangs:**

The comparative study of stability of various nanotubes shows remarkable difference in thermodynamic stability. We find that among 6-helix tubes, AT rich structure is thermodynamically most stable whereas for the 8-helix geometry original experimental sequence as well as the AT sequence are more stable. To study the effect of sequence on the thermodynamic stability of the DNA nanotube structures we have calculated the average potential energies. The potential energy of these nanostructures has been calculated using the per atom energy tool in LAMMPS[59] package. (**Fig. 9** and **Table3**) From our force-field calculations, we are able to extract the total energy of each atom interacting with the remainder of the DNA helix as well as interactions with all the ions and water molecules in the system. Once we have per atoms energies for each of the DNA nanotube atoms we sum them up to get the DNA nanotube energy. The energies are averaged over last 2 ns simulation data of all atoms of DNA nanotubes. The calculated potential energy of these nanostructures are given in Table 3 and has been shown in figure 8. To study the effect of overhangs, we have simulated AT rich and GC rich tubes without single stranded overhangs. In **Fig 10 and Fig. 11**, we compare the RMSD and radius for both the AT rich and GC rich 6-helix structures with and without overhangs. AT-rich structure with overhangs shows lesser RMSD implying better stability. This is also visible from the radius plot shown in **Fig 11 (b)** where structure with overhangs shows lesser fluctuation at both the ends and hence lower radius compared to the structures without overhangs. The unpaired hydrogen bonds in the single stranded are attributed to this extra stability. We have simulated the 6-helix structures with only AT and only GC composition as well. Here we see that AT structure has less RMSD and radius fluctuation, compared to GC structure implying higher thermodynamic stability of AT structure. (Supplementary information **Fig.S10**)

**Conclusion:**

We have presented an algorithm to generate a 3-d structure of various DNA nanotubes such as 6-helix nanotubes, 8-helix nanotubes and triangular tubes. Several nanosecond long MD simulations on these nanostructures provide critical information about microscopic structural feature and relative stability of these structures. AT rich and AT structures are more stable compared to other structure in the similar geometry. In particular we give an accurate estimate of the radius profile of these tubes which will be very important in the context of their cargo carrying application. Our simulation results also provide a direct estimate of the stretch modulus and persistence length of these nanotubes. Stretch moduli of the 6-helix nanotubes are in the range of 4000-4500 pN depending on the sequence. 8-helix nanotubes are distorted and have lower stretch modulus than 6-helix nanotubes. Hydrogen bond analysis and strain energy calculation demonstrates the relative stability of 6-helix nanotubes compared to the 8-helix geometry. The persistence lengths of these nanotubes are in the range of 6-7 μm and are in close agreement with the available experimental persistence length.[58] Whether our building methodology of forcefully fusing the helical domain of adjacent ds-DNA to create 8-helix structures led to a highly strained structure and results distortion or this behavior is inherent to 8-helix geometry needs further investigation.

## Acknowledgement

We thank Department of atomic energy (DAE) and DST India for financial assistance. We are grateful to Prof. Ned Seeman for valuable suggestions and critical reading of the manuscript. H. J. thanks CSIR for the research fellowship.

# Figures and Tables.

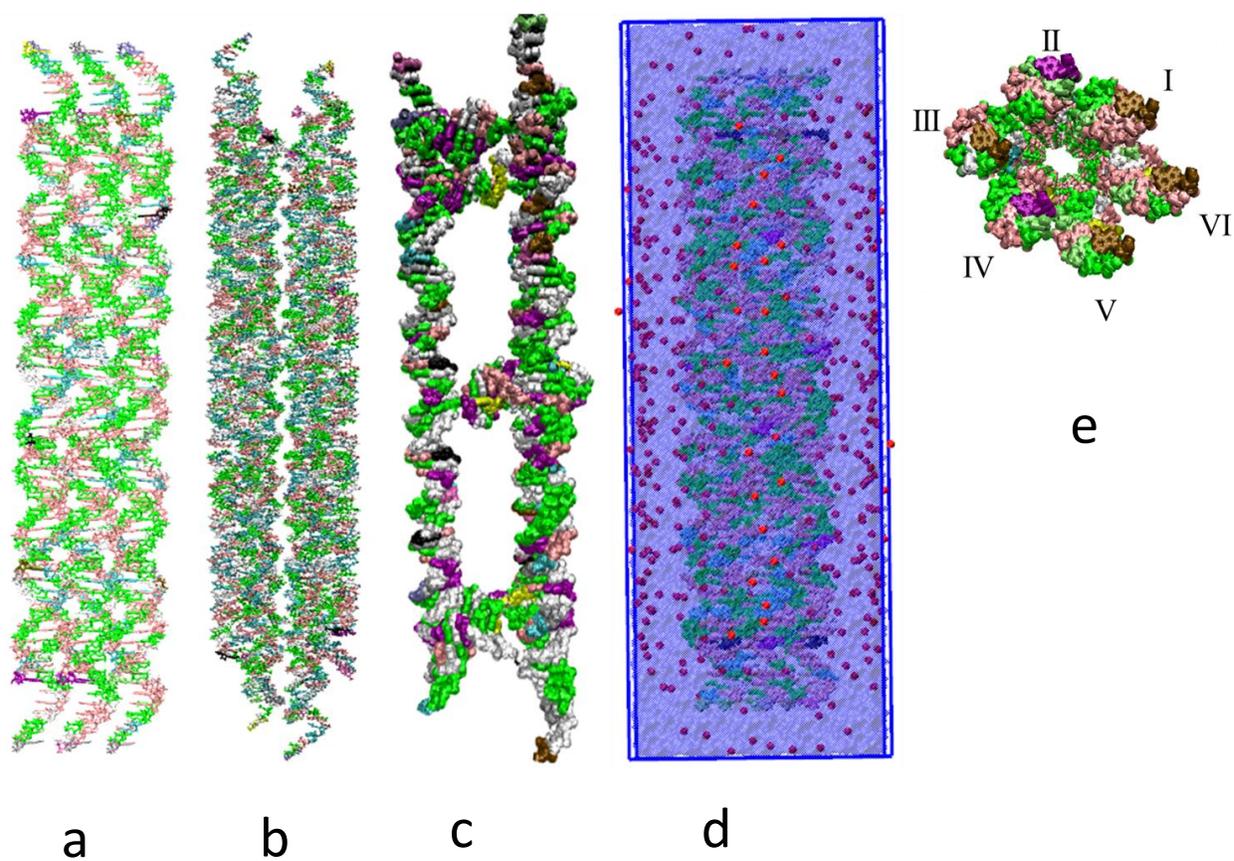

**Fig. 1**: Initial structure of various DNA Nanotubes built using in-house code. (a) 6-helix tube (b) 8- helix tube (c) 2 unit of triangular nanotube (d) 6-helix tube in explicit water and ions (e) top view of 6-helix nanotube with the helix identity.

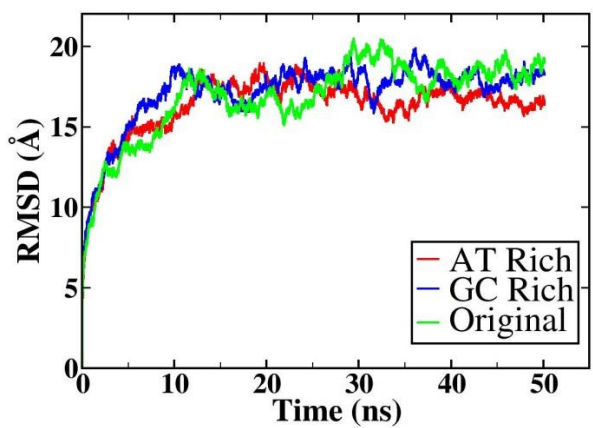 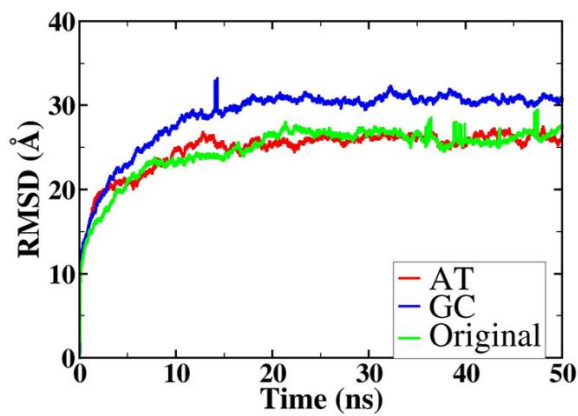

        a                                    b

**Fig. 2:** RMSD for various DNA nanotubes (a) 6-helix nanotubes and (b) 8-helix nanotubes. The RMSD is calculated with respect to the initial minimized structure. Original denotes the sequence used in original experimental paper. [19]

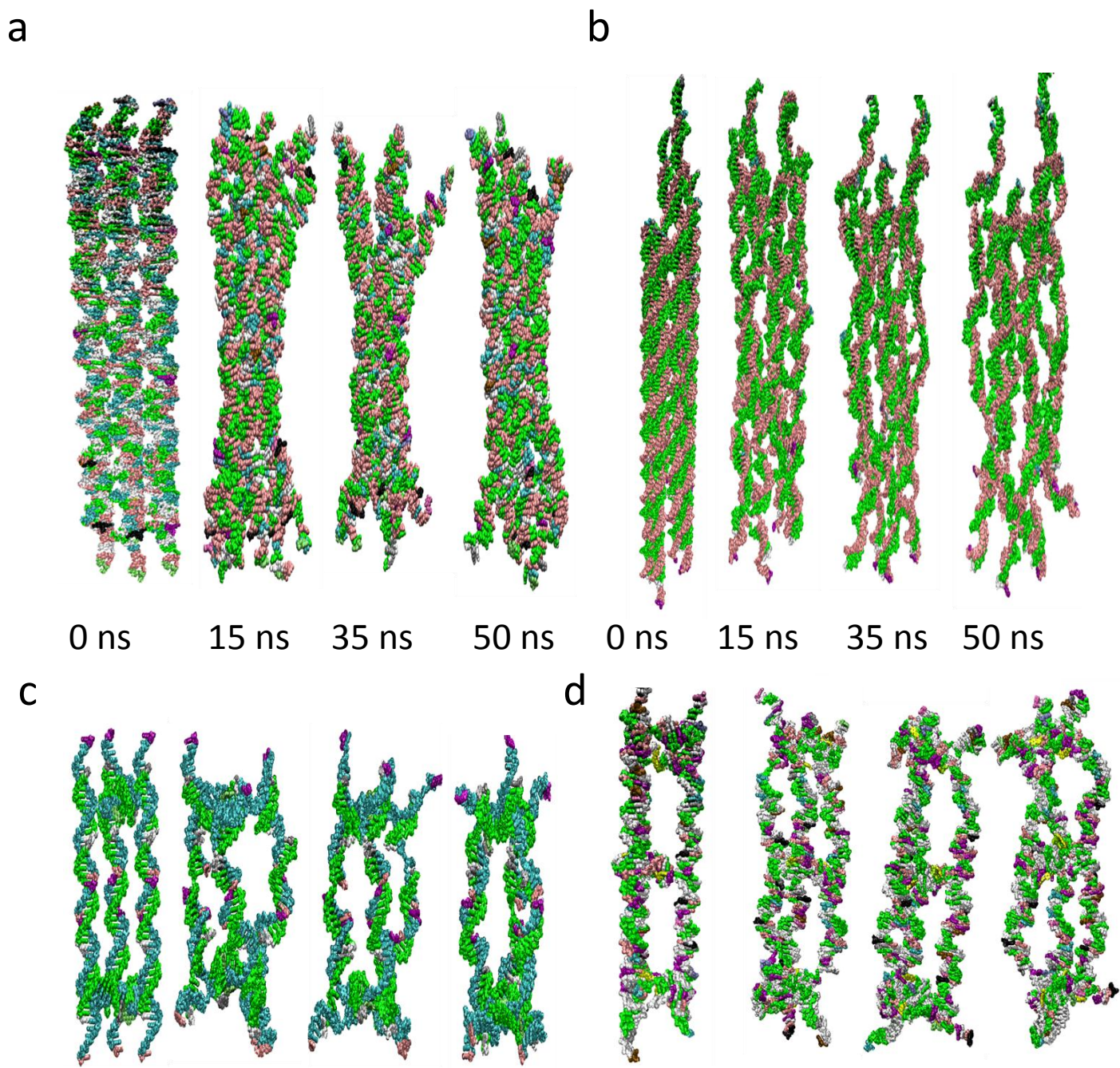

**Fig. 3:** Instantaneous snapshots of various nanotube structures at few ns intervals **(a)** 6-helix nanotube **(b)** 8- helix nanotube. Over nanosecond long time scale 8-helix nanotube loose it's tubular structure due to large strain in the topology. In contrast 6-helix nanotubes nicely maintain the tubular structure. Snapshots of **(c)** GC tri-tube structure and **(d)** two units of GC tri-tube. These tri-tube structures also maintain their shape during the simulation.

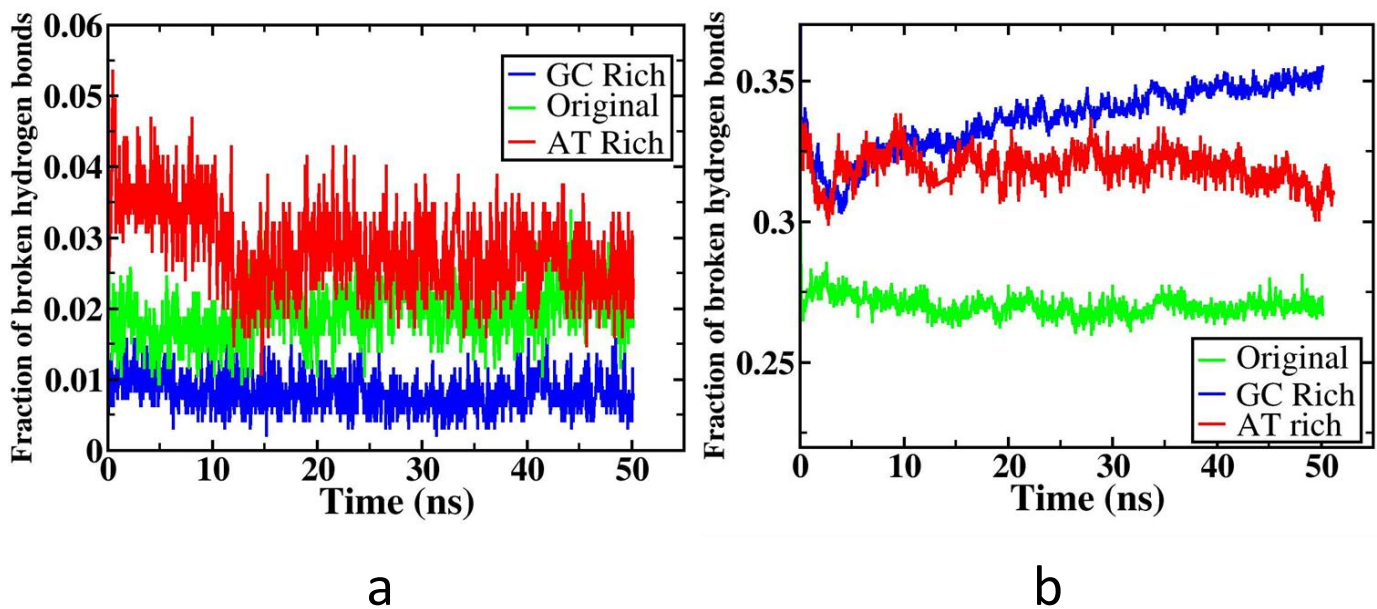

**Fig. 4:** Fraction of broken hydrogen bonds as a function of simulation time: **(a)** for 6-helix nanotubes **(b)** for 8-helix nanotubes. These plots show the better base stacking and higher stability for 6-helix DNA nanotubes.

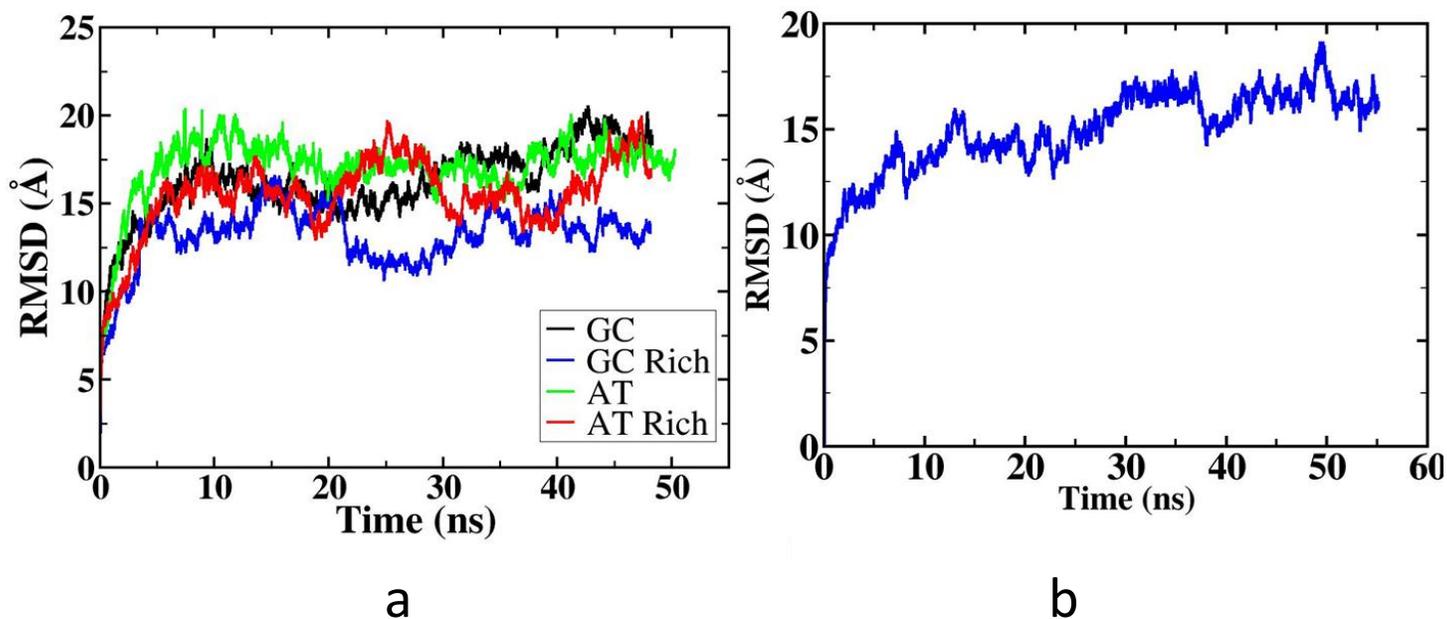

**Fig 5: (a)** RMSD for various triangular nanotubes having different base sequences. RMSD was calculated with respect to initial minimized structure. Note that the tri-tube RMSD is not very sensitive to the base composition. (b) RMSD for triangular tube of 2 units as shown in figure [1c]. For 2 repeat units the tri-tube structure is better stabilized and so the RMSD is smaller compared to the tri-tube with single unit.

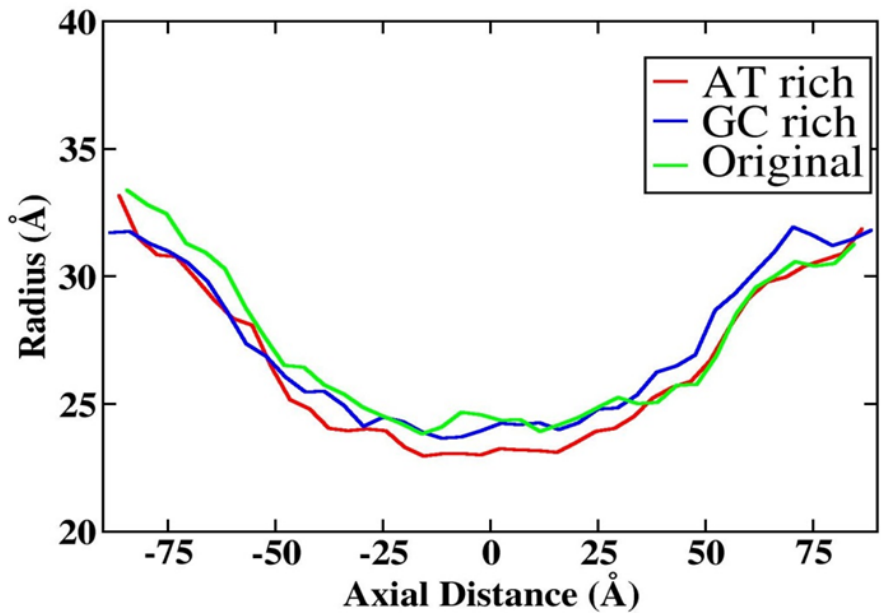

**Fig 6:** Radius profile of 6-helix DNA nanotubes for various base sequence composition. The middle of the bundle maintains nice tubular structure with a radius of 2.5 nm while both ends show larger radius because of end fraying. This is also visible from the snapshots shown in figure 3.

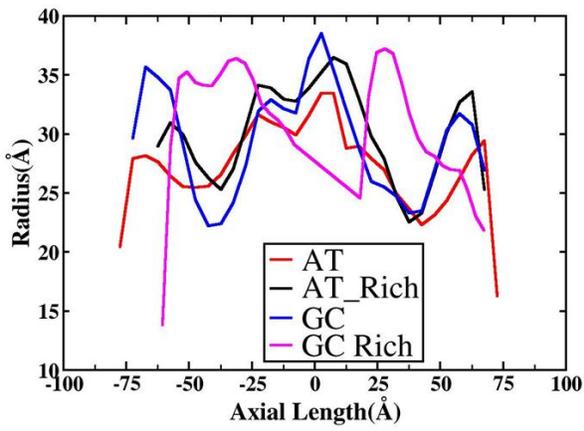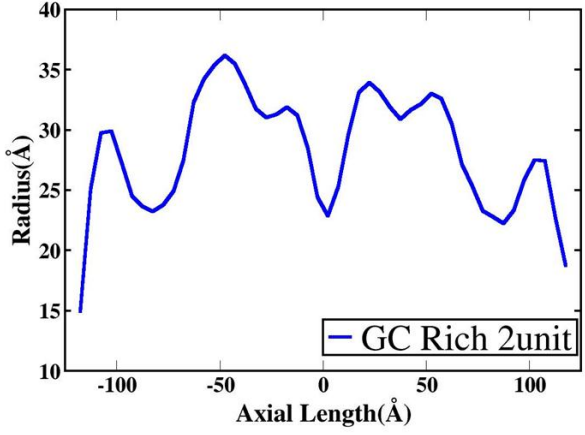

**Fig. 7**: Radius profile for various tri-tubes: (**a**) for single units, (**b**) for two units of tri-tube.

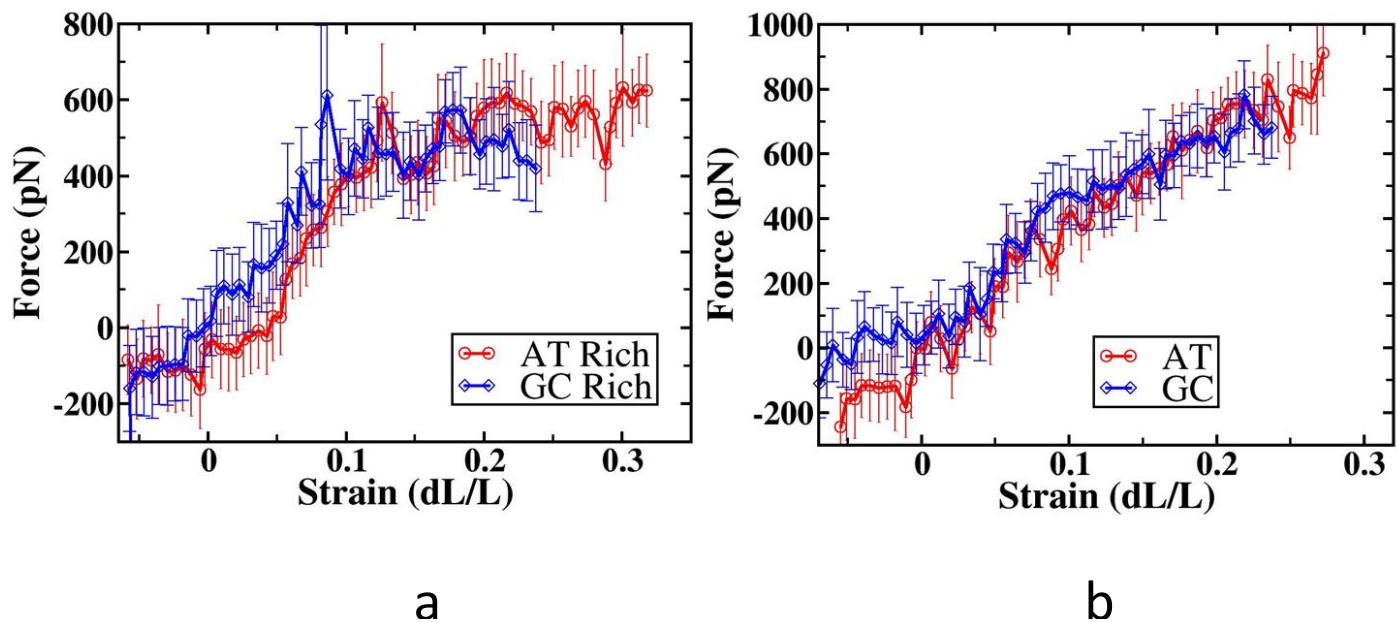

**Fig. 8:** Force-extension plot for (a) 6-helix and (b) 8-helix nanotubes for various base sequences obtained from steered molecular dynamics (SMD) simulation. From the linear part of the force extension relationship, we get the stretch modulus of the DNA nanotubes.

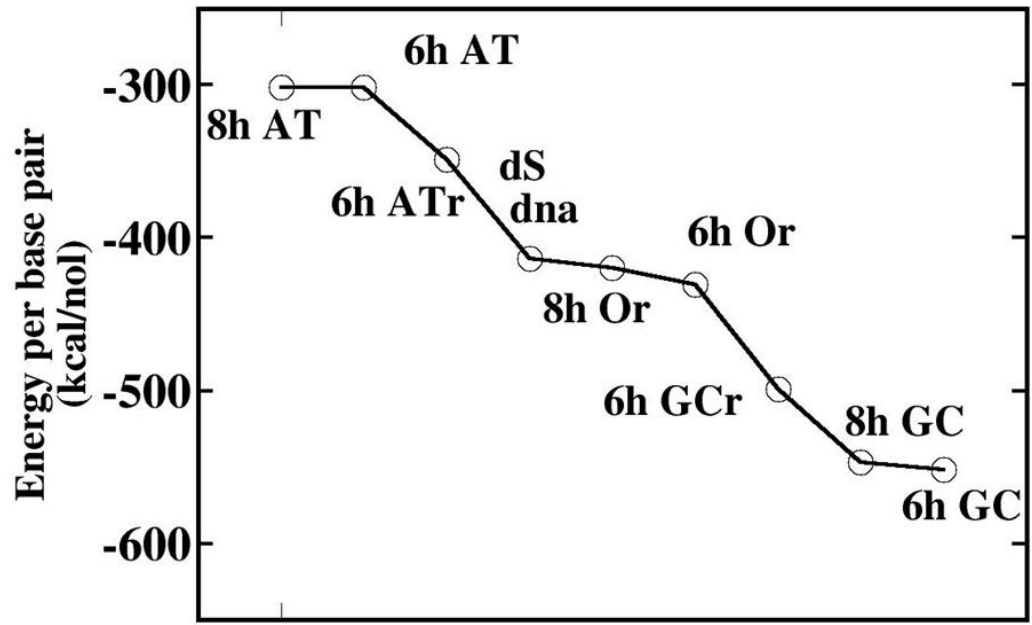

**Fig 9:** The potential energy profile of DNA nanostructures in various topologies.

The structures with higher GC contents are more stable on the basis of strain energies.

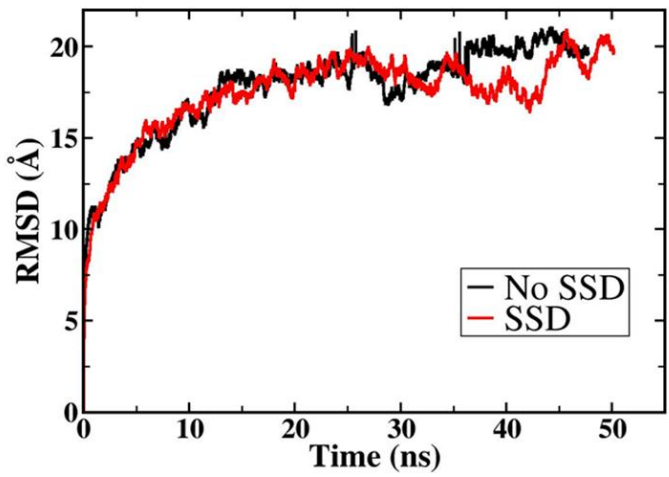 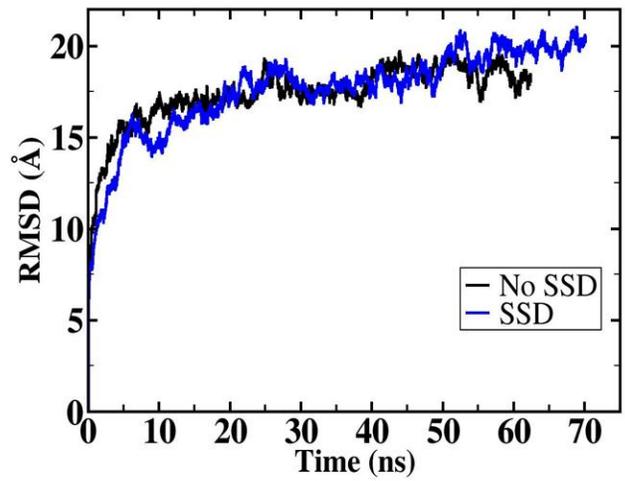

a   b

**Fig. 10**: RMSD comparison of 6-helix **(a)** AT rich and (b) GC rich structures with and without overhangs.

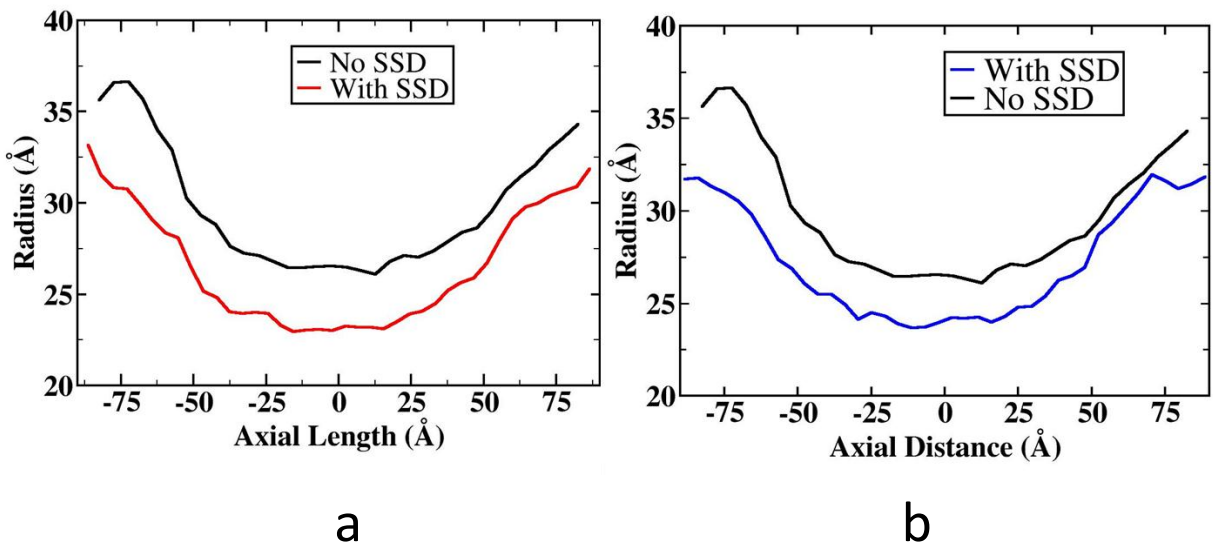

**Fig. 11:** Radius comparison of 6-helix **(a)** AT rich and **(b)** GC rich structures with and without overhangs. This indicates the better stability of structure with single stranded overhangs.

**Table 1: Details of the simulated system**

| System and Topology | No. of Atoms | Time (ns) | System and Topology | No. of Atoms | Time (ns) |
|---|---|---|---|---|---|
| 6-helix AT Rich | 268033 | 50 | 8-helix AT | 474244 | 50 |
| 6-helix GC Rich | 267203 | 50 | 8-helix GC | 473682 | 50 |
| 6-helix Original | 266654 | 50 | 8-helix Original | 475480 | 50 |
| 6- Helix AT | 265957 | 50 | Tritube AT | 180954 | 50 |
| 6- Helix GC | 265577 | 50 | Tritube GC | 183162 | 50 |
| 6-helix GC Rich (No SSD) | 209677 | 60 | Tritube GC Rich | 183712 | 50 |
| 6-helix AT Rich (No SSD) | 209007 | 50 | Tritube AT Rich | 191420 | 50 |
| 6-helix AT Rich (Pulling) | 309873 | 71 | 6-helix GC Rich (Pulling) | 307865 | 60 |
| 6-helix AT (Pulling) | 305379 | 72 | 6-helix GC (Pulling) | 312849 | 70 |
| 8-helix AT (Pulling) | 523229 | 63 | 8-helix GC (Pulling) | 521991 | 58 |

**Table 2: Radius, Stretch modulus and persistence length of various nanostructures studied.**

| Structure | Base pairs per helix /total | Radius (nm) | Length of Structure (nm) | Stretch Modulus (pN) | Persistence Length (nm) |
|---|---|---|---|---|---|
| **Double Helix** | 38/38 | 1 | 12.92 | 967 (± 58) | 58 |
| **6h AT rich** | 58/378 | 2.5 | 19.72 | 4269 (± 249) | 6425 |
| **6h GC rich** | 58/378 | 2.5 | 19.72 | 4468 (± 270) | 6724 |
| **6h AT** | 58/378 | 2.5 | 19.72 | 4397 (± 216) | 6617 |
| **6h GC** | 58/378 | 2.5 | 19.72 | 4507 (± 213) | 6783 |
| **8h AT** | 83/667 | 4.2 | 28.22 | 3826 (± 224) | 7460 |
| **8h GC** | 83/667 | 3.9 | 28.22 | 3899 (± 191) | 7658 |

**Table 3 : Energy of various Nanostructures**

| Structure | Number of Crossover and Base pairs | Potential Energy (using per atom energy break-up) (kcal/mol) | Standard Deviation | Energy per base-pair (kcal/mol) | Predicted using NN energy from ref ( kcal/mol) |
|---|---|---|---|---|---|
| dS DNA | 0, 12 | -4964.64 | ± 54.65 | -413.72 | -833.83 |
| 6-helix AT | 12,378 | -114113.30 | ± 305.49 | -301.88 | -24737.04 |
| 6-helix GC | 12,378 | -208518.80 | ±315.35 | -551.63 | -31000.88 |
| 6-helix ATr | 12, 378 | -131863.04 | ± 275.67 | -348.84 | -25788.37 |
| 6-helix GCr | 12, 378 | -188674.77 | ± 279.69 | -499.13 | -29415.76 |
| 6-helix original | 12, 378 | -162882.87 | ± 281.18 | -430.90 | -27728.90 |
| 8-helix AT | 14,667 | -201227.43 | ± 531.21 | -301.69 | -43709.94 |
| 8-helix GC | 14,667 | -364675.36 | ± 635.06 | -546.73 | -54748.37 |
| 8-helix Original | 14,667 | -280014.19 | ± 702.13 | -419.81 | -48562.12 |

**Supplementary Information**

**1. Construction of 6-helix and 8-helix DNA Nanotubes.**

To construct the 6-helix and 8 helix DNA nanotubes, we kept the double helix at the vertices of hexagon and octagon respectively. To arrange the helices into closed bundle, we fused the double helical arms of DNA with crossovers according to the closing geometrical angle of hexagon and octagon.

For example for a DNA with 10 base-pairs per turn, we can design the crossovers at 7 or 14 base- pairs spacing which will give us a closed angle of 120 °. In the case of 8 helix DNA nanotubes, we cannot get a perfectly closed regular octagon with geometrical angle 135° with DNA crossovers since it does not give any integer value of number of base pairs. The crossover for the corresponding geometry can happen either after a specified number of base pairs, or multiples of those to get a closed tube like structure. These spacing are as follows,

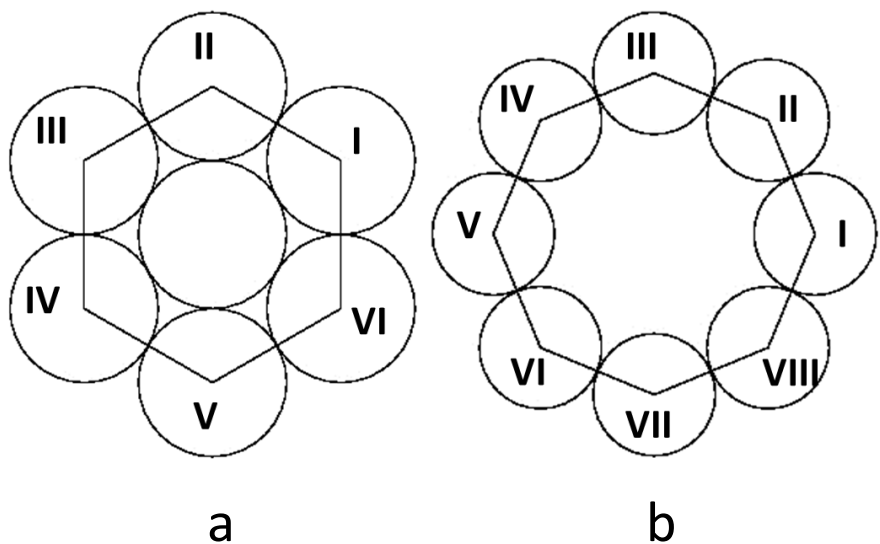

**Fig. S1:** Cross sectional view of **(a)** 6-helix and **(b)** 8-helix DNA Nanotubes. The numbers mentioned inside the circle are the helix identities.

To construct these nanostructures, we have designed a code using NAB module in AmberTools. This program takes the details about the structure from a sequence file. The sequence file is of a specific file format, which the program can read and use to create the structure. A portion of a sequence file is displayed below

```
cgactt   gatggagcaga|1|gctactt|2|ctacatc|  |gcattca|3|gtgctca|4|ggtacta|
------||ctacctgtct    cgatgaa    gatgtag    cgtaagt    cacgagt    ccatgat

cggtac   gtgacgatagg    acacatc    agatgtc    ttaggag    aggtcac    agtaacc
------||cactgctatcc|1|tgtgtag    tctacag|5|aatcctc|3|tccagtg    tcattgg
```

The first line represents the sense strand and the second line represents the antisense strand and so on. Together they form one double helix. An empty line is used as separators between individual double helices. Vertical bars are used to indicate breaks or nicks in the structure. A dash (-) is used when some regions are needed to be left single stranded. The number in between vertical bars serves as labels to make crossovers. A crossover is constructed between similarly labeled breaks. In the above example, crossovers are made between the 1 and 3 labeled breaks. The program generates the correct topology based on this input sequence file. To get the right orientations, specific NAB code has to be written in an orient function according to the requirements of the molecule. A generic orient function also exists which assumes a tubular structure for the entire molecule. It assumes that the molecule is made of parallel DNA double helices which have crossovers between them. The program takes as input a file containing the sequences and markers for the crossover point's locations. The program first reads the individual double helices and constructs broken helix structure for each of them. Next the individual broken helices are oriented about each other by the generic orient function. This is done in such a way as to minimize the root mean square distances between the atoms that need to be bonded. During the orientation process the double helices are only given three degrees of freedom to move, namely:

1. Rotation about their own helix axis.
2. Revolution about the helix axis of the molecule to which it is being bonded to.
3. Z-axis or vertical translation of the double helix such that all helix axes are always parallel.

The three parameters for the above three degrees of freedom are chosen by iterating over all possible values with appropriate step sizes and the values for which the RMSD is minimum, are chosen. Also the values are chosen such that there will not be any overlap with other helices and that they maintain a certain distance from each other. Next the individual bonds are made across the broken helices and the strands are merged so that the strand identity is preserved, as in the original molecule, so that the residues on each strand are correctly represented and numbered in the output. Finally, we get the PDB file of the structure for the AMBER MD simulation.

## 2. The connectivity of triangular DNA nanotubes: TBZ molecule

We have used the corner molecule to connect the triangular rungs of the triangular DNA nanotube. The molecule has been introduced to join the triangular rungs to the outer DNA forming the sculpture of triangular DNA nanotubes. The same molecule is used by Sleiman et. al[21], to construct the tri-tube geometry. This molecule has been designed with the xLeap module of AmberTools. GAFF[60] has been used to describe the interaction parameter for this molecule. Figure S2 shows the structure of TBZ molecule with GAFF atom type. Here is the picture of the corner molecule named as TBZ, (Fig. S2).

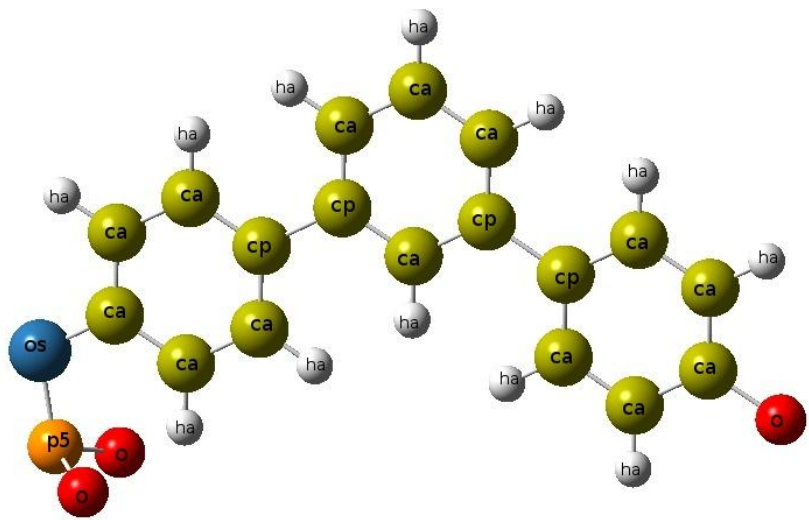

**Fig S2:** The connecting molecule TBZ for the triangular nanotube.

## 3. Snapshots of simulation.

The simulation without single stranded overhangs has also been done to compare the effect of overhangs. We have performed two sets of such systems, AT rich and GC rich nanotubes. The triangular nanotubes are also stable during the simulation. Here are some snapshots of the 6-helix nanotube as well as the triangular nanotubes during MD simulation.

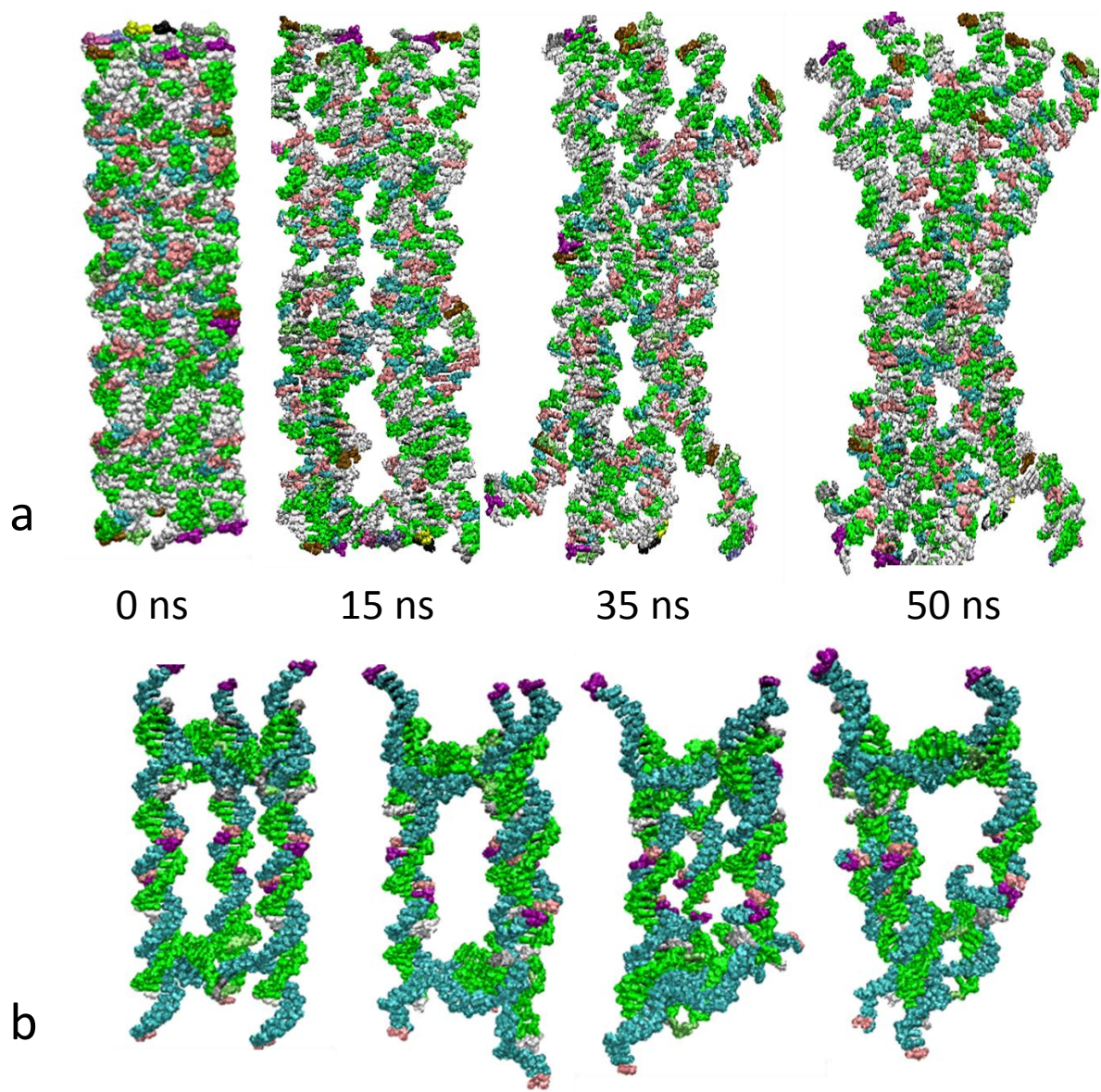

**Fig. S3:** Snapshots of simulation trajectories at various time steps,

**(a)** 6-helix DNA nanotube structure without Single stranded overhangs, **(b)** Triangular AT DNA nanotube during simulation.

## 4. RMSD with aqvist parameter.

The ion-water and ion-DNA interactions play an important role in the stability of DNA structure. So while studying the thermodynamic stability of DNA nanotubes with respect to various sequences; we have also simulated 6-helix and 8-helix structures with aqvist ion parameter for Na+ ion[38] and compare the results with those obtained using Joung and Cheatham ion parameters.[61] Structures simulated with aqvist ion parameter show higher RMSD compared with structures simulated using Joung and Cheatham parameters. Subsequently all the nanotubes structures reported in this paper have been simulated using Joung and Cheatham parameters. This implies the better suitability of Joung and Cheatham parameter for water-alkali ion-nucleic acid interactions.

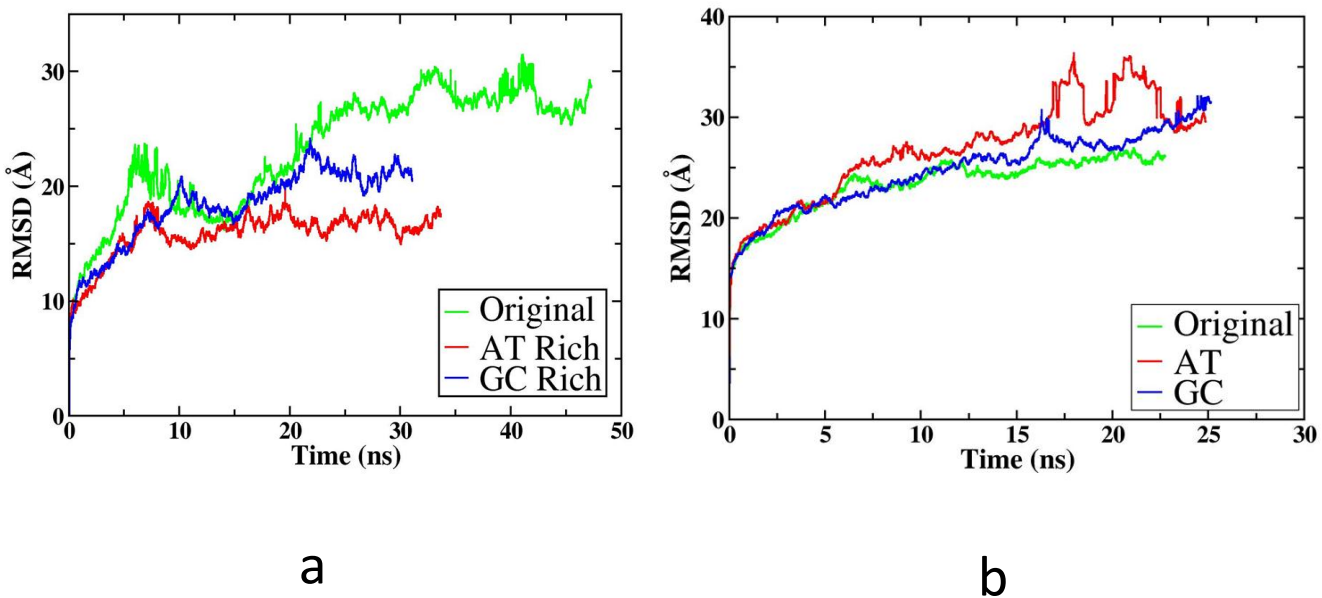

**Fig. S4:** RMSD for (a) 6-helix structures and (b) 8-helix structures with simulated using Aqvist ion parameter. The RMSD is calculated with respect to the initial minimized structure.

.

## 5. The 8 helix open structure.

Following the similar protocol used to build 6-helix DNA nanotubes, we put ds-DNA at the vertices of octagon in order to get the 8-helix DNA nanotube structure. But this protocol leads to a quit open structure, because of the geometry of B-DNA. Fig**. S5** shows the open helices of 8-helix structure. In order to get the closed tubular structure, we forcefully fused the helical domain of adjacent ds-DNA, which ultimately led to a highly strained structure. These structures try to minimize this dihedral strain during molecular dynamics simulation resulting in highly distorted tubular structures which can be easily seen in the instantaneous snapshots shown in **fig 3b**.

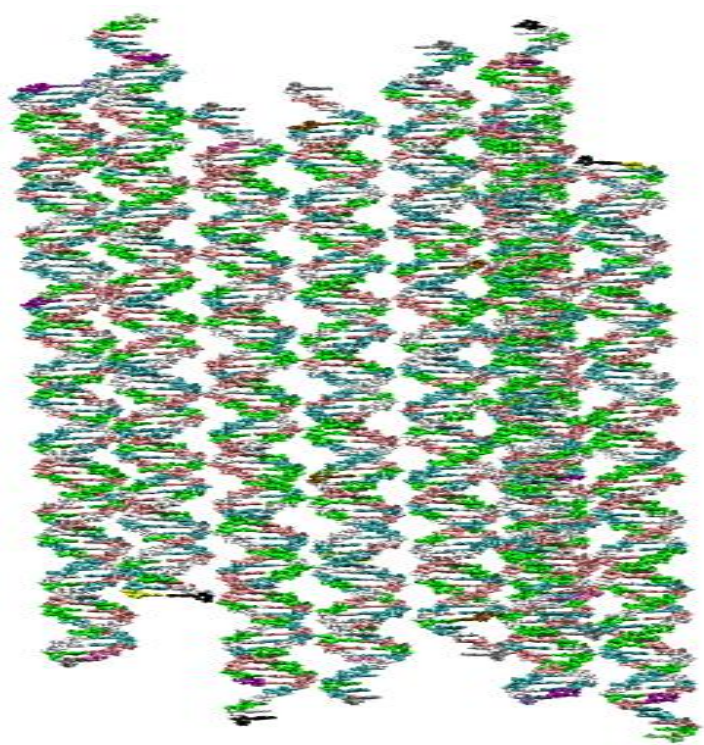

**Fig. S5:** The open structure of 8-helix bundle. Using the NAB code we make transformation and design crossovers to get a closed geometry.

## 6. Zigzag radius profile of 8-helix DNA nanotubes.

During several nanosecond long MD simulations, the 8-helix structures try to minimize the dihedral strain which results in highly deformed nanotube structure. This deformation gives rise to the erratic radius profile of all three 8-helix structures along the tube length. So overall, 8-helix structures are less stable due to its inherent closing angle which is not appropriate for crossover switching among helices.

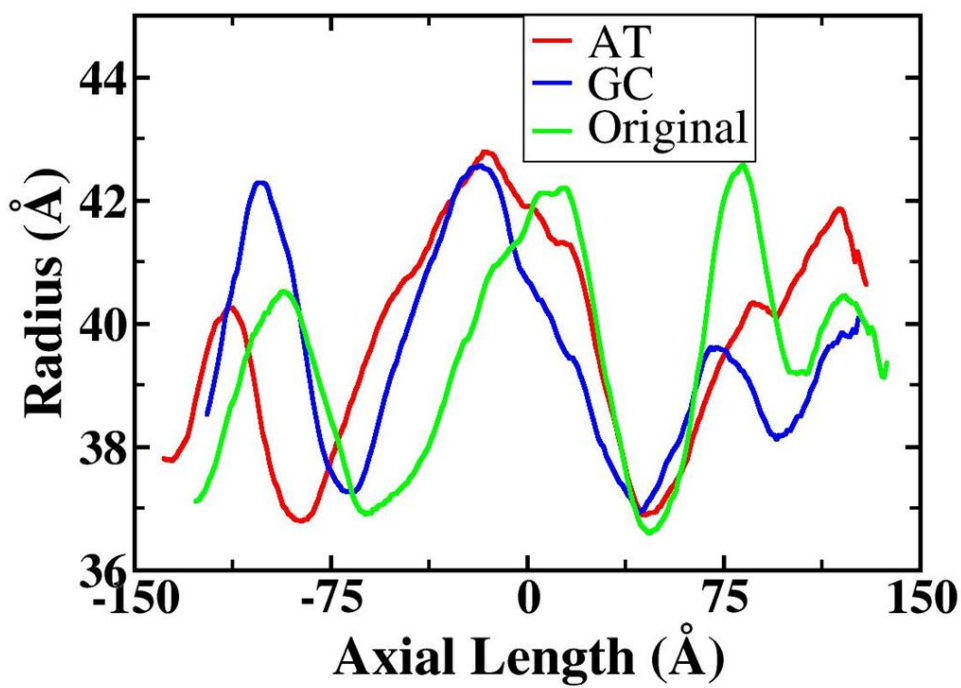

**Fig. S6:** The radius profile of 8-Helix DNA nanotubes with respect to the axial length.

## 7. Elastic properties of dS-DNA in constant velocity pulling simulation.

To explore the elastic response of DNA nanotubes, we pulled them in steered molecular dynamics (SMD) simulations in constant velocity ensemble. Note that the pulling rates in simulation are order of magnitude higher compared to the rates used in the experiments. Figure S7 shows the strain vs applied force (constant velocity ensemble) for 38-mer dS-DNA. From the linear region of this plot, we extract the value for stretch modulus for this structure.

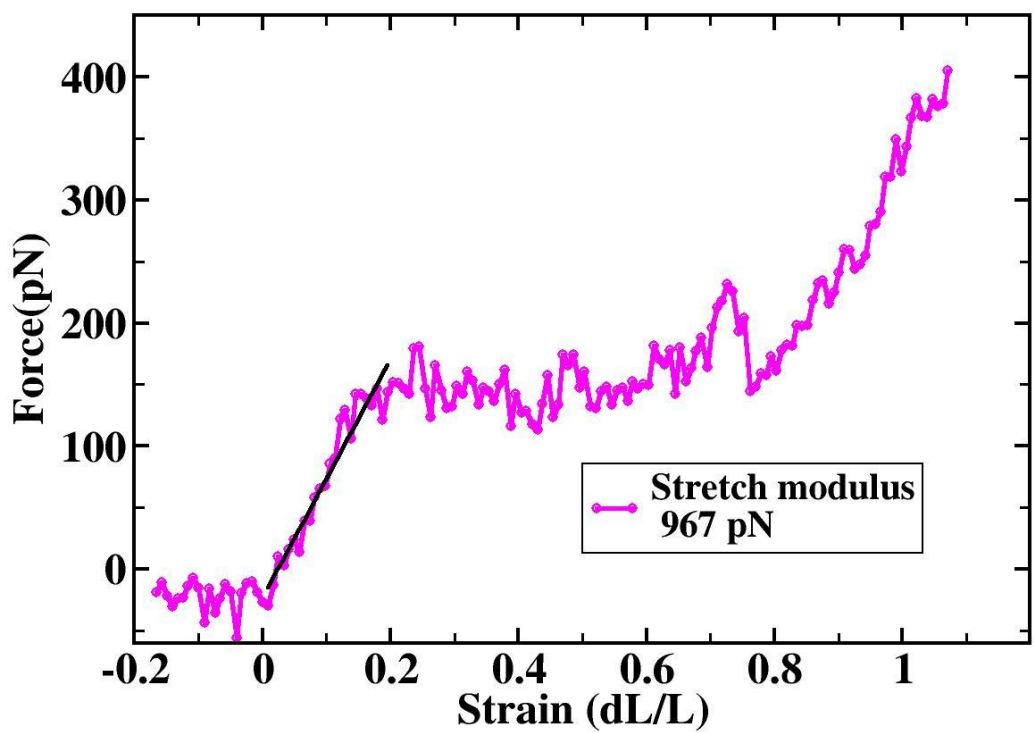

**Fig. S7**: The stress vs strain curve for 38-mer dS-DNA. The stretch modulus has been calculated from the linear region of the plot.

## 8. Snapshots of structures during constant velocity pulling.

Below we give instantaneous snapshots of the various nanotube structures at various strains during steered MD simulation. The tubes have been pulled from both the ends.

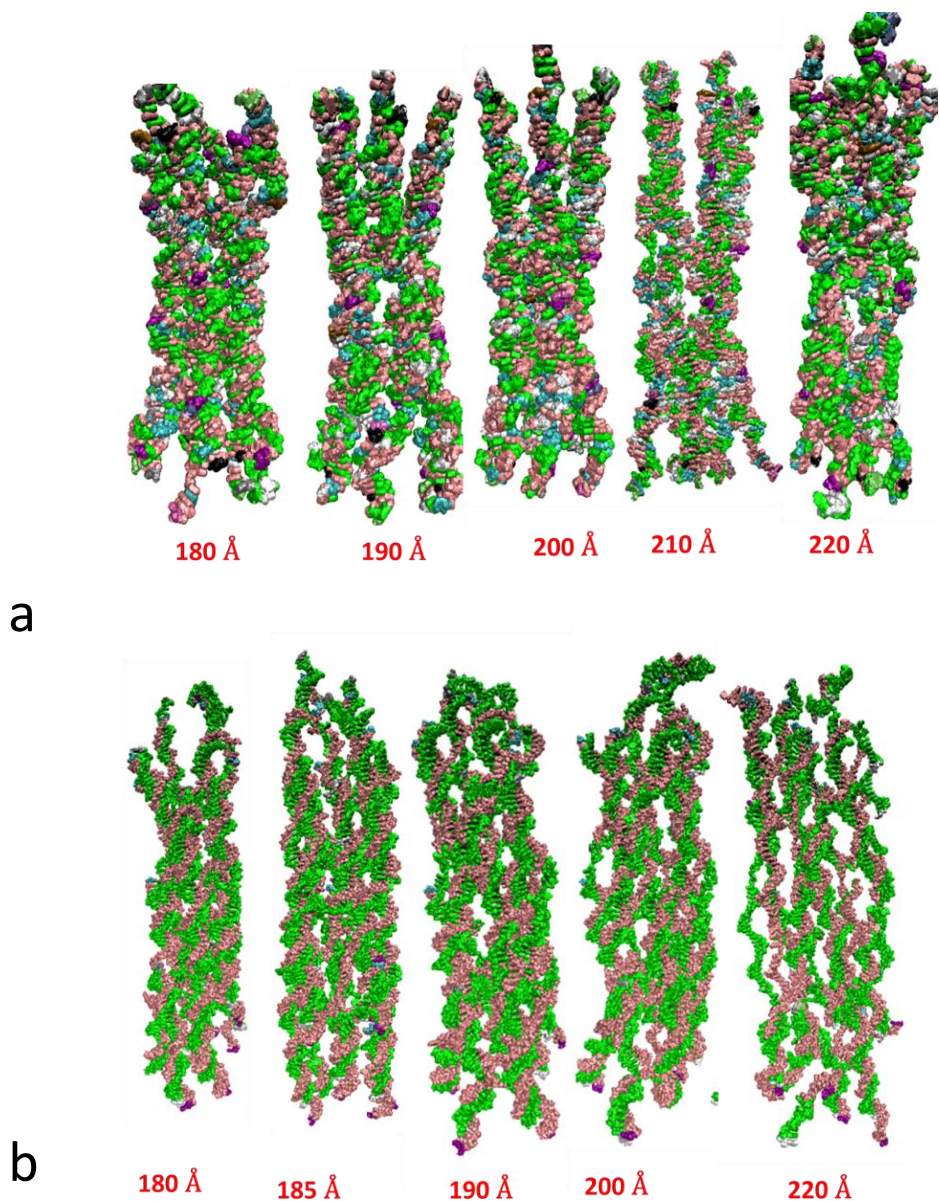

**Fig. S8 :** Snapshots of (**a**) 6-helix, (**b**) 8-helix DNA nanotube during SMD simulation.

## 9. Free energy calculation using WHAM analysis

From the SMD simulation we have calculated the free energy of the nanotube structure as a function of nanotube length using WHAM technique  The free energy as a function of the tube length for various nanotube geometry have been shown in figure S9. This gives us an estimate of the equilibrium length of these tubes.

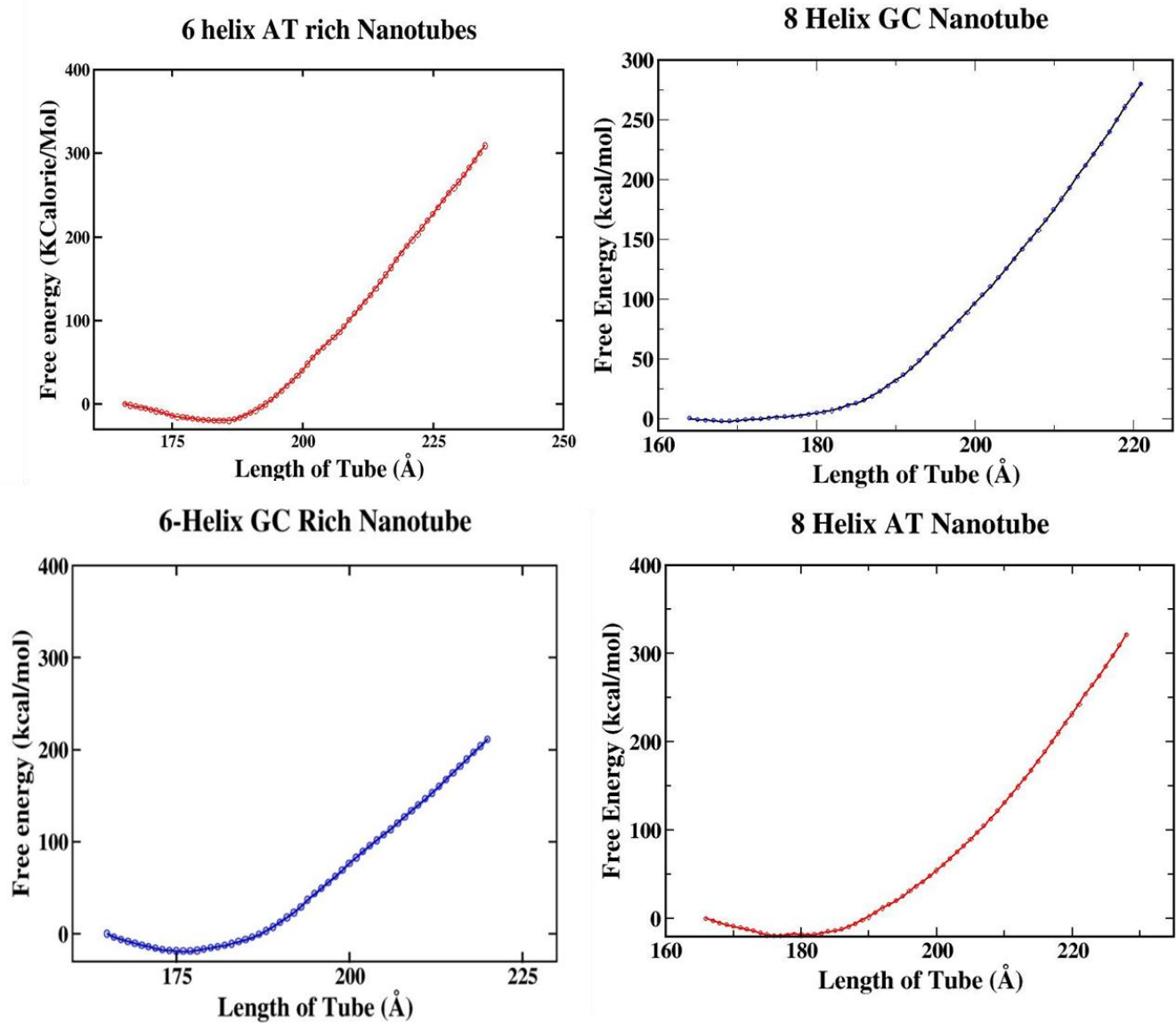

**Fig. S9:** Free energy for various DNA nanostructures as a function of the nanotube length obtained from the steered Molecular dynamics simulation. The minima of free energy plot correspond to the equilibrium length of these DNA nanotubes.

## 10. 6-helix DNA nanotubes made of pure AT and pure GC base composition.

To understand the effect of sequence on the stability of the 6-helix topology, we have done simulation of structures with only AT and only GC base sequences. Figure S10 (a) and (b) shows the RMSD and the radius profile for these structures. Structures with only AT base pairs are more stable. We have pulled these structures in constant velocity ensemble using steered molecular dynamics to calculate the stretch modulus of these structures. We see that the DNA nanotube composed of AT sequence is more stable which is clear from both RMSD and radius analysis. The stretch moduli of AT nanotube and GC nanotube are 4397 (± 216) pN and 4507 (± 213) pN respectively.

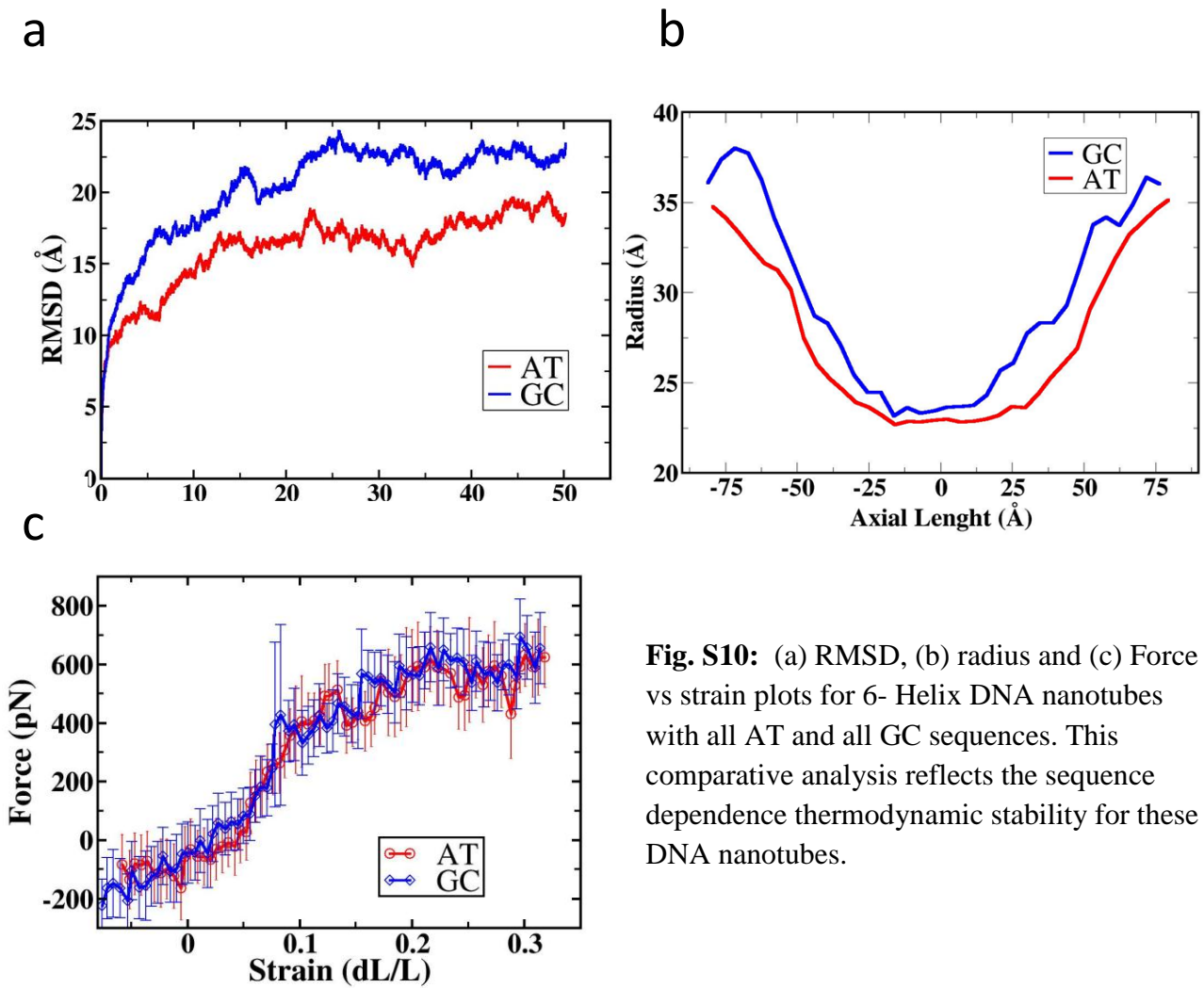

**Fig. S10:** (a) RMSD, (b) radius and (c) Force vs strain plots for 6- Helix DNA nanotubes with all AT and all GC sequences. This comparative analysis reflects the sequence dependence thermodynamic stability for these DNA nanotubes.

**11. Effect of pulling velocities on DNA Nanotubes.**

6-helix AT rich structure has been pulled with three different velocities to explore the role of pulling velocities to the elastic response under constant velocity SMD simulation. We see that the slope of the linear region of force vs strain plot is almost similar with respect to all three pulling velocities. As expected, the pulling force required is less with low pulling velocity compared to high pulling velocity for the same strain in the structure. The plateau region begins at lower forces for slow pulling velocities. While pulling with 0.05 m/s, the simulation is more realistic but it is computationally expensive.

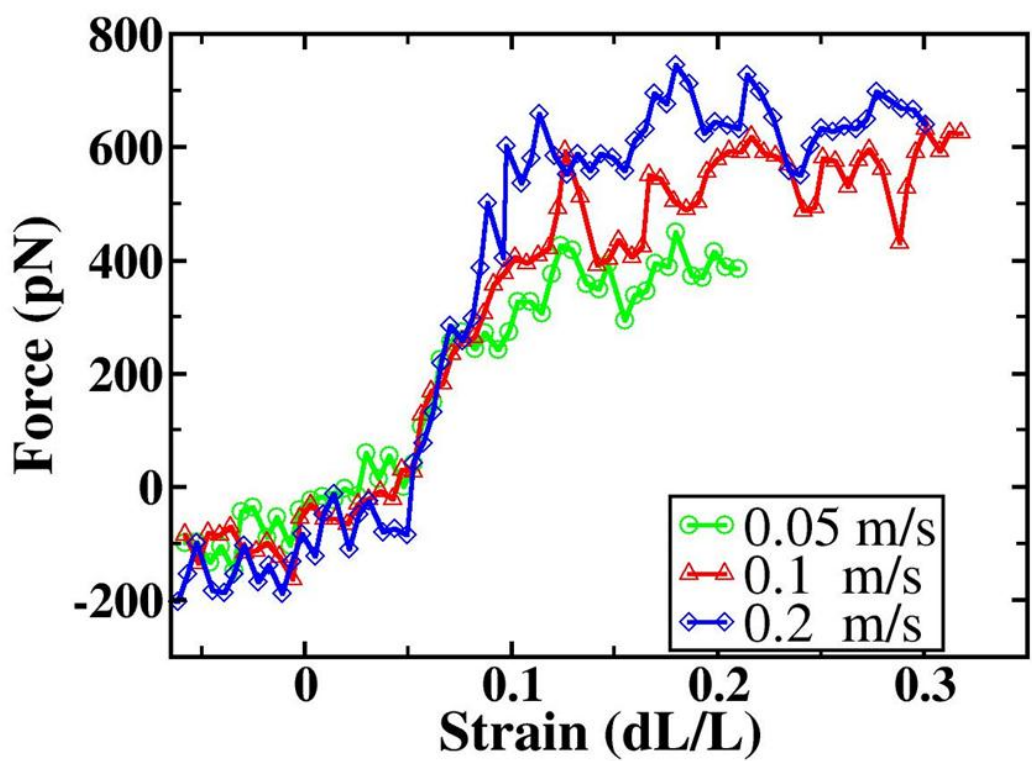

**Fig. S11:** Force vs Strain plot for 6-helix AT rich structure at different pulling velocities. The plateau region approaches to the smaller force values as we decrease the pulling velocity.

## 12. Snapshots of the cross sectional view from the top of the nanotubes

This figure shows the variation of the cross section of DNA nanotubes tubes with respect to the simulation time.

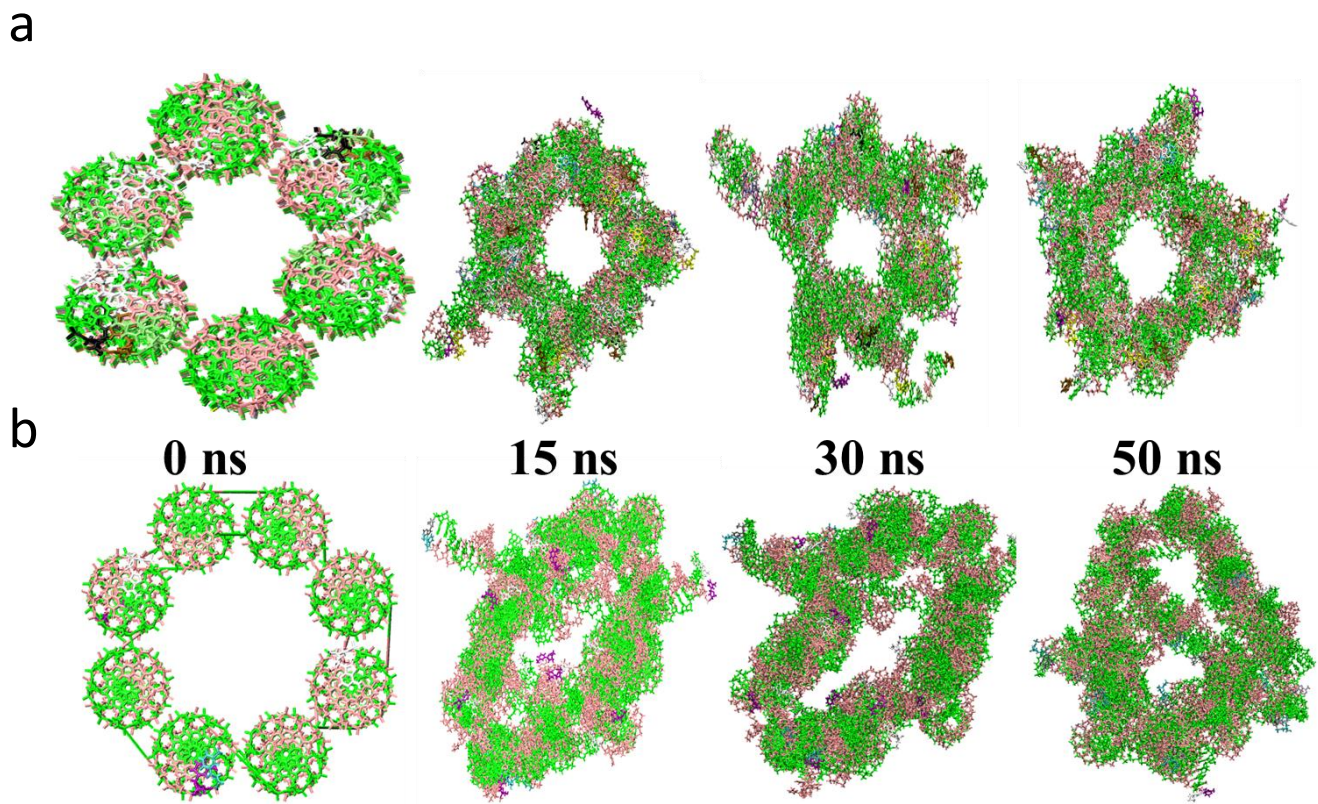

**Fig. S12**. The cross sectional view of DNA nanotubes at various times steps during the simulation. (a) 6 helix DNA nanotubes. (b) 8 helix DNA nanotubes.